# Updating Weight Values for Function Point Counting

Wei Xia[1], Danny Ho[2], Luiz Fernando Capretz[3, *], Faheem Ahmed[4]
[1]HSBC Bank Canada, IT Department, Vancouver, Canada
[2]NFA Estimation Inc., London, Canada
[3]University of Western Ontario, Dept. Electrical & Computer Engineering, London, Canada
[4]United Arab Emirates University, College of Information Technology, Al-Ain, U.A.E

*corresponding author – lcapretz@eng.uwo.ca

**Abstract**

While software development productivity has grown rapidly, the weight values assigned to count standard Function Point (FP) created at IBM twenty-five years ago have never been updated. This obsolescence raises critical questions about the validity of the weight values; it also creates other problems such as ambiguous classification, crisp boundary, as well as subjective and locally defined weight values. All of these challenges reveal the need to calibrate FP in order to reflect both the specific software application context and the trend of today's software development techniques more accurately. We have created a FP calibration model that incorporates the learning ability of neural networks as well as the capability of capturing human knowledge using fuzzy logic. The empirical validation using ISBSG Data Repository (release 8) shows an average improvement of 22% in the accuracy of software effort estimations with the new calibration.

**Index Terms**

Function point analysis, software size measure, effort prediction model, software estimation

1. INTRODUCTION

Accurate software estimation is a crucial part of any software project so that the project can be priced adequately and resources allocated appropriately. If the management underestimates the actual cost, the organization can lose money on the project, or even worse an inaccurate estimation can ruin a small software company. Conversely, if the management overestimates, the client may decide that, on the basis of cost-benefit analysis or return of investment, there is no point in having the software built. Alternatively, the client may contract another company whose estimate is more reasonable. Additionally, the client certainly wants to know when the project will be delivered, if the management is unable to keep to its schedule, then at best the organization loses credibility, at worst penalty clauses are invoked. In all cases, the managers responsible for the software estimation have a lot of explaining to do. Hence it is clear that accurate software estimation is vital.

Therefore, many models for estimating software development effort have been proposed: COCOMO [1, 2], SLIM [3], and Function Point (FP) [4] are arguably the most popular. These models can be considered algorithmic models, that is, pre-specified formulas for estimating development efforts that are calibrated from historical data. COCOMO and SLIM assume Source Lines of Code (SLOC) as a major input. FP, however, is based on higher-level features of a software project, such as the size of files, the types of transactions, and the number of reports; these features facilitate estimation early in the software life cycle.

Algorithmic effort prediction models are limited by their inability to cope with vagueness and imprecision in the early stages of the software life cycle. Although software engineering has been influenced by a number of ideas from different fields [5], software estimation models combining algorithmic models with machine learning approaches, such as neural networks and fuzzy logic, have been viewed with scepticism by the majority of software managers [6].

Srinivasan and Fisher [7] illustrate a neural network learning approach to estimating software development effort known as *Back-propagation*. They indicate possible advantages of the approach relative to traditional models, but also point out limitations that motivate continued research. Furthermore, MacDonell [8] also considers the applicability of fuzzy logic modelling methods to the task of software source code sizing, and suggests that fuzzy predictive models can outperform their traditional regression-based counterparts, particularly with refinement using data and knowledge. Ahmed, Saliu and AlGhamdi [9] discuss an adaptive fuzzy logic framework for software effort prediction that incorporates experts' knowledge; they demonstrate the capabilities of the framework through empirical validation carried out on artificial data sets and on the COCOMO public database of completed projects. Xu and Khoshgoftaar [10] present a fuzzy identification cost estimation modeling technique to deal with linguistic data and automatically generate fuzzy membership functions and rules; they report that their model provides a method of cost estimation that is significantly better than COCOMO.

Briefly, neural network techniques are based on the principle of learning from historical data, whereas fuzzy logic is a method used to make rational decisions in an environment of uncertainty and vagueness. However, fuzzy logic alone does not enable learning from the historical database of software projects. Once the concept of fuzzy logic is incorporated into the neural network, the result is a neuro-fuzzy system that combines the advantages of both techniques.

Neuro-fuzzy cost estimation models are more appropriate when uncertainties and imprecise information are accounted for. Huang *et al*. [11] present a general framework that combines neuro-fuzzy techniques with algorithmic approaches, and show that such a combination can result in an average improvement of 15% for software estimation accuracy. Huang *et al*. [12] used a neuro-fuzzy approach as a front-end to COCOMO, this combination yielded better estimation results than COCOMO alone. The research presented in this paper extends that general framework to FP counting method.

FP is a metric of measuring software size that was first proposed by Albrecht [4] at IBM in 1979. The introduction of FP represented a major step in comparison with SLOC counting, because FP focuses on system "functionality" rather than on calculating SLOC. Based on the FP theory, other variations, such as International FP Users Group (IFPUG) [13], COSMIC [14], and Mark II [15] were created.

Research on the combination of the machine learning approach with FP was conducted by Finnie, Wittig and Desharnais [16], who compared three estimation techniques using FP as an estimate of system size. The models considered are based on regression analysis, artificial neural networks and case-based reasoning. Although regression models performed poorly on the given data set, the authors observed that both artificial neural networks and case-based reasoning appeared to be valuable for software estimation models. Hence, they concluded that case-based reasoning is appealing because of its similarity to the expert judgement approach and for its potential in supporting human judgement.

Additionally, Yau and Tsoi [17] introduced a fuzzified FP analysis model to help software size estimators express their judgment and use the fuzzy B-spline membership function to derive their assessment values. Their weakness involved the used of limited in-house software, which significantly limits the validation of their model. Lima, Farias and Belchior [18] also proposed the application of concepts and properties from the fuzzy set theory in order to perform fuzzy FP analysis; a prototype that automates the calculation of FPs using the fuzzified model was created. However, as in the case of Yau and Tsoi [17], the calibration was done using a small database comprised of legacy systems, developed mainly in Natural-2, Microsoft Access and Microsoft Visual Basic, a fact which limits the generality of their work.

A new FP weight system using an artificial neural network was established by Al-Hajri *et al*. [19]. Their research was similar to ours in the fact that they also used the data set provided by the International Software Benchmarking Standards Group (ISBSG) in order to validate their model. In their work, they replaced the original complexity table with a new table gathered with the training methods from neural networks. Although their results are quite accurate, the correlation is still unsatisfactory due to the wide variation of data points with many outliers.

Previous research projects have indicated that the combination of machine learning approaches and algorithmic models yields a more accurate prediction of software costs and effort, which is competitive with traditional algorithmic estimators. However, our proposed neuro-fuzzy model goes even further: it is a unique combination of statistical analysis, neural networks and fuzzy logic. Specifically, we obtained an equation from statistical analysis, defined a suite of fuzzy sets to represent human judgement, and used a neural network to learn from a comprehensive historical database of software projects. First, we will briefly introduce the theory of function points, and then in the next section, we will illustrate its weaknesses and limitations.

FP Analysis is a process used to calculate software functional size. Currently, the most pervasive version is regulated in the Counting Practices Manual - version 4.2, which was released by the International Function Point User Group (IFPUG) [13]. Counting FP requires the identification of five types of functional components: Internal Logical Files (ILF), External Interface Files (EIF), External Inputs (EI), External Outputs (EO) and External Inquiries (EQ). Each functional component is classified as a certain complexity based on its associated file numbers such as Data Element Types (DET), File Types Referenced (FTR) and Record Element Types (RET). The complexity matrices for the five components are shown in Table I. Table II illustrates how each function component is then assigned a weight according to its complexity. The Unadjusted Function Point (UFP) is calculated with Equation 1, where $W_{ij}$ are the complexity weights and $Z_{ij}$ are the counts for each function component.

$$UFP = \sum_{i=1}^{5} \sum_{j=1}^{3} Z_{ij} \cdot W_{ij} \qquad \text{Equation 1}$$

**Table I - COMPLEXITY MATRIX FOR FP FUNCTION COMPONENTS**

| ILF/EIF | DET | | | EI | DET | | | EO/EQ | DET | | |
|---|---|---|---|---|---|---|---|---|---|---|---|
| RET | 1-19 | 20-50 | 51+ | FTR | 1-4 | 5-15 | 16+ | FTR | 1-5 | 6-19 | 20+ |
| 1 | Low | Low | Avg | 0-1 | Low | Low | Avg | 0-1 | Low | Low | Avg |
| 2-5 | Low | Avg | High | 2 | Low | Avg | High | 2-3 | Low | Avg | High |
| 6+ | Avg | High | High | 3+ | Avg | High | High | 4+ | Avg | High | High |

**Table II - FUNCTION COMPONENT COMPLEXITY WEIGHT ASSIGNMENT**

| Component | Low | Average | High |
|---|---|---|---|
| External Inputs | 3 | 4 | 6 |
| External Outputs | 4 | 5 | 7 |
| External Inquiries | 3 | 4 | 6 |
| Internal Logical Files | 7 | 10 | 15 |
| External Interface Files | 5 | 7 | 10 |

Once calculated, UFP is multiplied by a Value Adjustment Factor (VAF), which takes into account the supposed contribution of technical and quality requirements. The VAF is calculated from 14 General System Characteristics (GSC), using Equation 2; the GSC includes the characteristics used to evaluate the overall complexity of the software.

$$VAF = 0.65 + 0.01\sum_{i=1}^{14} Ci \qquad \text{Equation 2}$$

where *Ci* is the Degree of Influence (DI) rating of each GSC.

Finally, an FP is calculated by the multiplication of UFP and VAF, as expressed in Equation 3.

$$FP = UFP \times VAF \qquad \text{Equation 3}$$

Despite its well recognized usefulness as software metric, FP has its own drawbacks and weaknesses. The next section expands on these difficulties, particularly the problems with the current FP complexity weight systems. Section 3 proposes a FP calibration model, termed the neuro-fuzzy FP (NFFP) model, which overcomes those problems. In section 4 we describe the experimental methodology and discuss the results of performance evaluation, the reliability and validity of the proposed model. Finally, the last section summarizes the conclusions of this work.

## 2. PROBLEMS WITH THE FP COMPLEXITY WEIGHT SYSTEM

The FP complexity weight system refers to all the complexity calculations and weight values expressed in FP. Five problems with the FP complexity weight system are identified. Problems 1 and 2 are rooted in the classification of the UFP, whereas problems 3, 4 and 5 arise from the weight values of UFP.

### 2.1. Problem 1: Ambiguous Complexity Classification

Each of the five FP function components (ILF, EIF, EI, EO and EQ) is classified as low, average or high, according to the complexity matrices that are based on the counts of each component's associated files, such as DET, RET and FTR. Such complexity classification is easy to perform, but the categorization itself can be ambiguous. For example, Table III shows a software project with two ILF, A and B. Both A and B have 3 RET, but A has 50 DET while B has 20 DET. According to the complexity matrix, both A and B are classified as having the same complexity and are assigned the identical weight value of 10. However, A has 30 more DET than B and is therefore more complex. Despite their difference, they are assigned the same complexity weight, thus exposing the problem of ambiguous classification.

*2.2. Problem 2: Crisp Boundary in Complexity Classification*

The boundary between two different complexity classifications is very crisp. An example is given in Table III, where one software project has two ILF, B and C. Both B and C have 3 RET, but B has 20 DET, while C has 19 DET. B is classified as average by the complexity matrix and assigned a weight of 10, whereas C is classified as low and assigned a weight of 7. B has only one more DET than C and they both have the same number of RET. However, B has been classified as average and assigned three more weight units than C, because the boundary between B and C is very crisp with no smooth transition between the values.

**Table III - PROBLEM 1(AMBIGUOUS CLASSIFICATION) AND PROBLEM 2 (CRISP BOUNDARY)**

|  | ILF A | ILF B | ILF C |
|---|---|---|---|
| **DET** | 50 | 20 | 19 |
| **RET** | 3 | 3 | 3 |
| **Complexity Classification** | Average | Average | Low |
| **Weight Value** | 10 | 10 | 7 |
| **Problem 1** | Ambiguous Classification | | |
| **Problem 2** | | Crisp Boundary | |

*2.3. Problem 3: Weight Values Obsolete*

The weight values of unadjusted FP [4] are said to reflect the size of the software. They have never been updated since being introduced in 1979 and are applied universally. In contrast, software development methods have evolved steadily since 1979, but today's software differs drastically from what it was over two decades ago. Such an imbalance prompts the question: "Are these weight values are still applicable to today's software or whether are they obsolete?"

*2.4. Problem 4: Weight Values Defined Subjectively*

The weight values of unadjusted FP were determined by Albrecht by "debate and trial" [4], based on his experience and knowledge. Albrecht contributed significantly to the theory of FP; nevertheless, with no actual follow-up projects to justify his values, the question remains as to whether the weight values were defined subjectively without convincing support or whether they reflect objective data.

*2.5. Problem 5: Weight Values Defined Locally*

The weight values of unadjusted FP were decided based on the study of data processing systems at IBM. The assignment of weight values was restricted to only one organization and to only a limited amount of software types; however, this set of weight values is applied universally and is not limited to one organization or one type of software. Thus, one cannot be sure if weight values that were defined locally can be applied on a global basis.

*2.6. Remarks on the Encountered Problems*

The existing weight system of FP does not measure complexity perfectly. Problems 1 and 2 do not fully reflect software complexity under a specific application context; they definitely need calibration. Problems 3, 4 and 5 do not reflect the trends of today's software and need further calibration as well. In an attempt to address these problems, this paper proposes a novel FP calibration model.

### 3. FP CALIBRATION MODEL USING NEURO-FUZZY

In introducing the new FP calibration model, called a neuro-fuzzy FP (NFFP) model, we have three main objectives: to improve the FP complexity weight system, to calibrate the weight values of the unadjusted FP and to produce a calibrated FP count for more accurate measurements.

*3.1. Technical View of the Model*

The neuro-fuzzy FP model is a unique approach that incorporates FP measurements with the neural networks, fuzzy logic and statistical regression techniques; a technical view of this model is depicted in Figure 1. The first component, statistical regression, is a mathematical technique used to represent the relationship between selected values and observed values from the statistical data. Secondly, the neural network technique is based on the principle of learning from previous data. This neural network is trained with a series of inputs and desired outputs from the training data so as to minimize the prediction error [20]. Once the training is complete and the appropriate weights for the network links are determined, new inputs are presented to the neural network to predict the corresponding estimation of the response variable.

The final component of our model, fuzzy logic, is a technique used to make rational decisions in an environment of uncertainty and imprecision [21], [22]. It is rich in its capability to represent the human linguistic ability with the terms of fuzzy set, fuzzy membership function, fuzzy rules, and the fuzzy inference process [23]. Once the concept of fuzzy logic is incorporated into the neural network, the result is a neuro-fuzzy system that combines the advantages of both techniques [24].

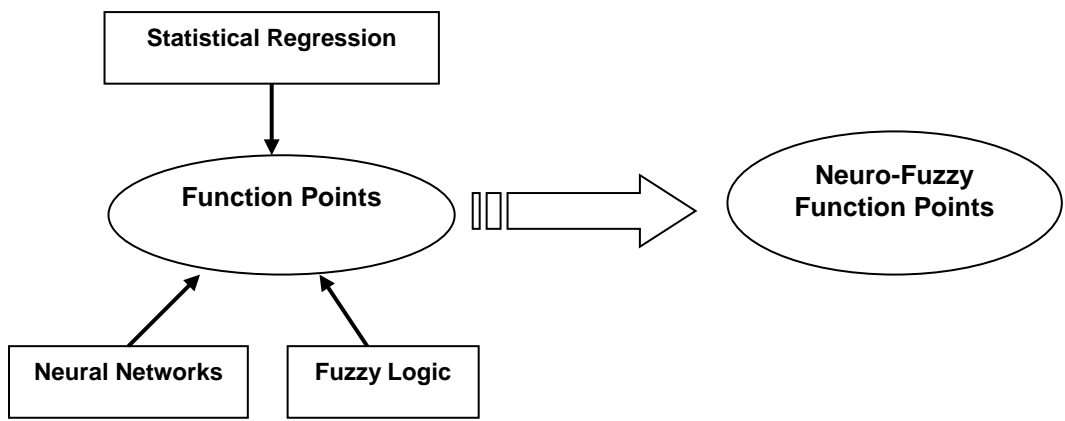

Figure 1 - Overview of the FP Calibration Model

*3.2. Layered Architecture*

The neuro-fuzzy FP model consists of three layers: the input, processing and output layers. A block diagram of this model is illustrated in Figure 2.

*a) Input layer:* The input layer of the model consists of the pre-defined fuzzy complexity measurement system based on experts' experience, which is subjective knowledge, and the project data provided by ISBSG [25], considered to be objective knowledge. The pre-defined system produces an exact complexity degree for each function component of FP. The project data imported in this layer is used to extract an estimation equation by means of the statistical regression technique and to train the neural network in the processing layer.

*b) Processing layer:* An equation that estimates work effort is derived in the processing layer by analyzing the project data imported from the input layer using a statistical regression technique. The equation is then applied in the neural network as an activation function. Also, the neural network learning block calibrates the weight values of unadjusted FP by learning from the historical project data.

c) *Output layer:* The calibrated weight values of unadjusted FP (UFP) are generated from the neural-network learning block. They are used in the output layer to adjust the parameters of the pre-defined fuzzy complexity measurement system, producing an adjusted fuzzy complexity measurement system. The pre-defined and the post-tuned (adjusted) fuzzy measurement system combine to form the fuzzy logic component of the model.

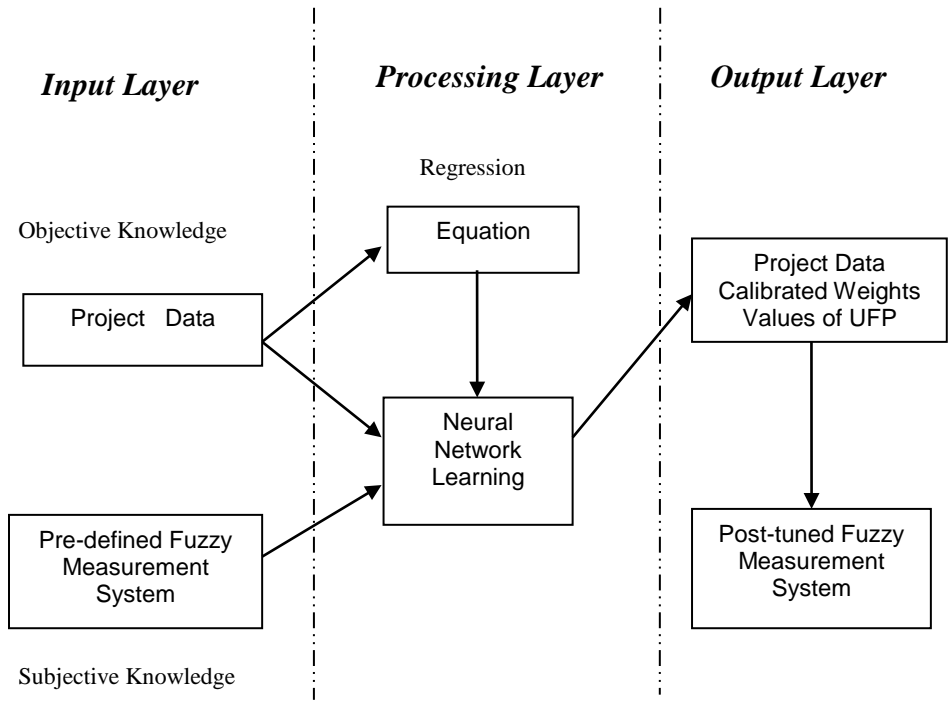

**Figure 2 - Layered Architecture of FP Calibration Model**

*3.3. Fuzzy Complexity Measurement System*

The fuzzy part of the neuro-fuzzy FP model is composed of two fuzzy complexity measurement systems: the pre-defined system in the input layer and the adjusted system in the output layer. The two fuzzy systems share the same structure, but the pre-defined system uses the original unadjusted FP weight values whereas the adjusted system uses the calibrated weight values. Using fuzzy logic in a comprehensive way produces an exact complexity degree of each FP component with the associated file numbers and overcomes Problems 1 and 2 by obtaining a more precise categorization.

As mentioned in the first section, the five FP function components are classified according to the complexity matrices. The inputs in the original weight matrices are the numbers of files associated with each function component, and the output is the component's complexity classification: *low*, *average* or *high*.

We define three new linguistic terms: *small*, *medium* and *large*, to express the inputs qualitatively. For example, if an ILF has one RET, we assume that the ILF's RET input is small. Similarly, if an EI has six DETs, we assume that EI's DET input is medium. Also, we use the linguistic terms *low*, *average* and *high* for the output, which is the same as the original matrices. For example, the linguistic terms of low, average, and high are used to describe the weight values of 7, 10 and 15 of ILF respectively. Thus, the linguistic terms defined are consistent with the original complexity matrices. We unified all of the complexity matrices for the five FP components to one equivalent linguistic complexity matrix shown in Table IV.

Table IV - LINGUISTIC COMPLEXITY MATRIX

|  | Input 1 | | |
| --- | --- | --- | --- |
| Input 2 | Small | Medium | Large |
| Small | Low | Low | Average |
| Medium | Low | Average | High |
| Large | Average | High | High |

The fuzzy sets are defined to represent the linguistic terms in the complexity matrix. Each input and output in the linguistic complexity matrix is represented by a fuzzy set named after its linguistic term. The membership grade is captured through the membership functions of each fuzzy set. There are several basic types of fuzzy membership functions [24]; the trapezoidal and the triangular types of membership functions are selected because the complexity increases linearly with the file numbers and also because these types of membership functions are appropriate in preserving the values. The inputs are of the trapezoidal type and the outputs are of the triangular type. Since all of the original complexity matrices are now unified to the linguistic complexity matrix (Table IV), all the inputs and outputs of the five FP function components can be represented by fuzzy sets. The five components share the same fuzzy set structure but have different parameters in fuzzy membership functions; the parameters are assigned to make the boundary more gradual according to the original complexity matrices.

An example of the complexity matrix for ILF/EIF is given to illustrate the fuzzy set structure and the definition of parameters. As illustrated in Figure 3(a), the inputs of DET and RET are represented in the fuzzy sets of *small*, *medium*, and *large*, and Table V shows the parameters of the membership functions. The parameters *a*, *b*, *c*, and *d* ($a<b\leq c<d$) determine the coordinates of the trapezoid, where *a* and *d* locate the "feet" of the trapezoid, and *b* and *c* locate the "shoulders". Figure 3(b) shows the outputs of the complexity matrix of ILF/EIF that are represented in the fuzzy sets of *low*, *average*, and *high,* and Table V demonstrates the parameters of the fuzzy membership functions. The parameters *a*, *b*, and *c* ($a<b<c$) define the coordinates of the triangle, where *a* and *c* locate the "feet" of the triangle, and *b* locates the "peak". Using a similar method, we can define the fuzzy sets for the remaining function components (EI, EO and EQ) and fuzzify all of the inputs and outputs in the FP complexity weight matrices.

The fuzzy inference process using the Mamdani approach [26] is applied based on the fuzzy sets and rules. For each fuzzy rule, we apply the minimum method to evaluate the "AND" operation and obtain the antecedent result as a single number. The antecedent of each rule implies the consequence using the minimum method. All of the consequences are aggregated to an output fuzzy set using the maximum method. Eventually, the output fuzzy set is defuzzified to a single number using the centroid calculation method, which takes the center of gravity of the final fuzzy space and produces a clear output value.

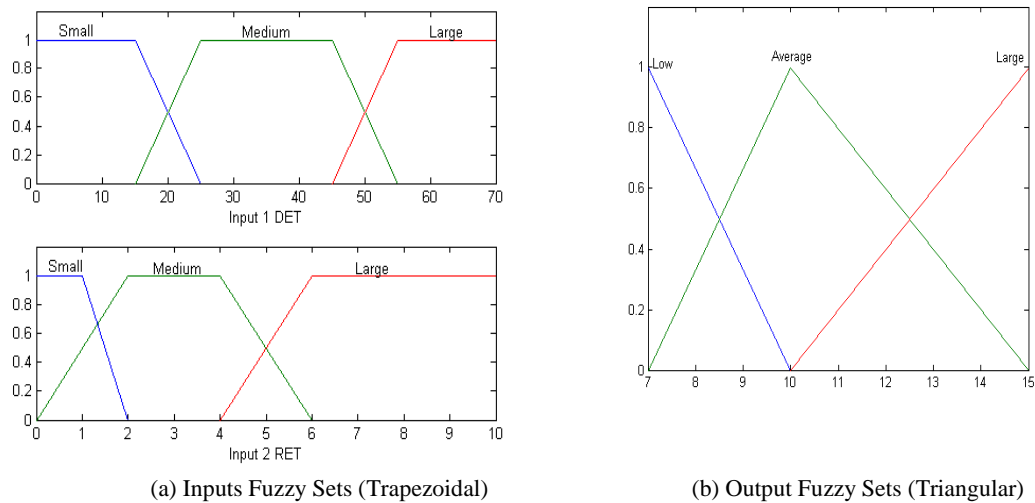

(a) Inputs Fuzzy Sets (Trapezoidal)   (b) Output Fuzzy Sets (Triangular)

**Figure 3 - Neuro-Fuzzy FP Model Fuzzy Sets**

**Table V - NEURO-FUZZY FP MODEL FUZZY SETS PARAMETERS**

| Input1 | Small | Medium | Large | Input2 | Small | Medium | Large | Output | Low | Average | High |
|---|---|---|---|---|---|---|---|---|---|---|---|
| a | 0 | 15 | 45 | a | 0 | 0 | 4 | A | 7 | 7 | 10 |
| b | 0 | 25 | 55 | b | 0 | 2 | 6 | B | 7 | 10 | 15 |
| c | 15 | 45 | 70 | c | 1 | 4 | 10 | C | 10 | 15 | 15 |
| d | 25 | 55 | 70 | d | 2 | 6 | 10 | | | | |

*3.3.1. Problems Tackled with the Proposed Approach*

If we apply our method to the previous examples illustrating the FP complexity Problems 1 and 2, we obtain three weight values that are much more accurate. Table VI shows the original weight values and the fuzzy weight values of ILF A, B and C, therefore demonstrating that the new fuzzy weight values solve both Problem 1 (ambiguous classification) and Problem 2 (crisp boundary).

**Table VI - PROBLEMS SOLVED USING FUZZY LOGIC**

|  | ILF A | ILF B | ILF C |
|---|---|---|---|
| **DET** | 50 | 20 | 19 |
| **RET** | 3 | 3 | 3 |
| **Original Weight Value** | 10 | 10 | 7 |
| **Problem 1** | Ambiguous Classification | | |
| **Problem 2** | | Crisp Boundary | |
| **Fuzzy Weight Value** | 11.4 | 10.4 | 10.2 |

*3.4. Extraction of the Estimation Equation*

The aim of this section is to establish an equation that can estimate the software cost in terms of work effort and then be used in the neural network training as an activation function. Although similar estimation equations are proposed in [27] and [28], we decided to create a new equation based on the ISBSG Data Repository - release 8.

*3.4.1. Data Preparation*

In order to obtain a reasonable equation, the raw data must be filtered by several criteria, since it is necessary to ensure that the model is built on the basis of a reliable data set.

According to the information provided by the ISBSG [29], only the data points whose *quality* ratings are A or B should be considered. Hence, we discarded the C and D rating projects and were left with the remaining 1,445 projects.

Next, we wanted to ensure that further analysis is based on the most widely used counting methods. Although FP has several variations of counting methods, including IFPUG [13], COSMIC [14] and Mark II [15], the IFPUG method is the most popular standard of counting as it is used by 90% of the projects (1,827 out of 2,027) in the whole ISBSG data set. Hence, only the projects counted using the IFPUG method were selected.

The work efforts of the projects are recorded at different *resource levels*, including level one (development team), level two (development support team), level three (computer operations) and level four (end user or client). To ensure the reliability of further analysis, we ensured that our data was based on the same resource level as the majority of projects. Thus, we chose to have the projects recorded at the first resource level, a level which covers 70% (1,433 out of 2,027) of the projects in the whole ISBSG data set.

Development type is the final criterion on which we based our analysis. There are three major *development types* of the projects in the ISBSG data set: new development (838 projects), enhancement (1,132 projects), and redevelopment (55 projects). However, there is one utility development project and one purchase package project that do not belong to any one of the three major development types. The new development and redevelopment projects can be classified into one large group, whose results are calculated using the original FP equation (Equation 3) of the first section. However, the enhancement projects are calculated using a very different equation [13] (see Equation 4); these projects should be treated as a separate case study.

$$FPenhance = (UFPadd + UFPchange) \times VAFafter + UFPdelete \times VAFbefore \quad \text{Equation 4}$$

The new development and redevelopment projects that were based on the original FP equation were considered. In order for the weight values to be further calibrated using the neural network, the neuro-fuzzy FP model considers the projects that offer the 15 categories of unadjusted FP, in other words, the five function components with low, average and high classifications. Furthermore, in order to add flexibility to the model, we chose those projects that provided the 14 GSC rating values; almost all of the projects that provide the 15 UFP categories provide 14 GSC rating values as well.

The application of all of these criteria results in a significant decrease in the number of projects. As a result, a subset of 409 projects was obtained where the quality rating is A or B, the counting method is IFPUG, the effort resource level is one, and the development type is new or redevelopment. A further reduction of the projects to those that provide the 15 unadjusted FP categories and 14 GSC rating values resulted in a 184 project data set that was used to build the equation. Similarly, Angelis, Stamelos and Morisio [30] conducted research on the ISBSG data repository and encountered the same problem; they used ISBSG - release 6, which contains 789 projects, but had only 63 projects left after applying the filtering criteria.

*3.4.2. Statistical Analysis*

After applying the filtering criteria, the data obtained was transformed to satisfy the assumptions for regression analysis; the positive relationship between the work effort and size has been reported in [27], [28], and [31]. FP is a functional software size that is calculated by multiplying the UFP and the VAF, as shown in Equation 3. However, the definition of VAF, which is supposed to reflect the technical complexity, has been criticized for overlapping and for being outdated [30], [32]. The unadjusted FP has been standardized as an unadjusted functional size measurement through the International Organization for Standardization (ISO) [33], but the VAF has not. The ISBSG - release 8 data field description document recommends using the field of "Normalized Work Effort" as a project work effort. Thus, we have chosen UFP as the software size, and the normalized work effort as the effort in statistical analysis.

The statistical regression analysis assumes that the underlying data are normally distributed; however, our original data is highly skewed. To approximate a normal distribution, we apply a logarithmic transformation to these variables in order to decrease the large values and to bring the data closer together. After the transformation, *ln* UFP and *ln* Work Effort are approximately normally distributed. The relationship between the work effort and the size is illustrated, using two-dimensional graphs as shown in Figures 4(a) and 4(b), before and after the logarithmic transformation respectively. Furthermore, there is an obviously positive linear relationship between effort and size after the logarithmic transformation.

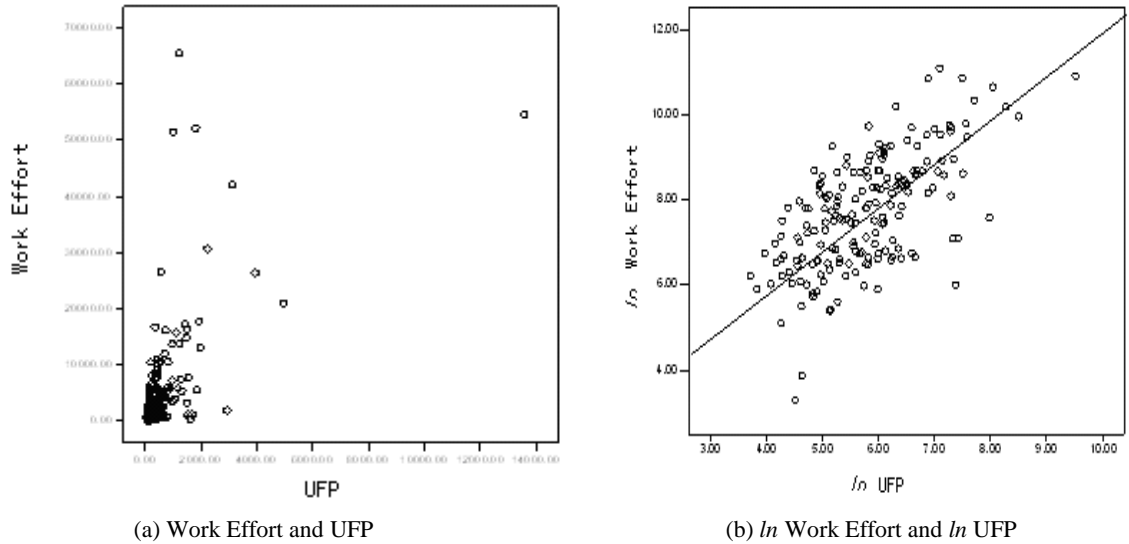

(a) Work Effort and UFP                    (b) *ln* Work Effort and *ln* UFP

**Figure 4 - Graphs of UFP and Work Effort**

The regression process applied to the data set is automated by the statistical software package SPSS version 12. An equation in the form of Equation 5 was calculated, and its equivalent form is shown in Equation 6:

$$ln\ Effort = \alpha \cdot ln\ UFP + \beta \qquad \text{Equation 5}$$

$$Effort = A \cdot UFP^B \qquad \text{Equation 6}$$

where $\alpha, \beta, A, B$ are all coefficients calculated from the regression process.

Certain post regression analysis was done to check the validity of the regression. It is observed that the residuals are normally distributed, independent, with a constant variance and having a mean value that equals zero. Therefore, the assumptions for statistical regression analysis are satisfied and we can conclude that Equation 5 and its equivalent form Equation 6 are valid.

*3.4.3 Remarks on Equation Extraction*

Though of simple form, Equation 6 is derived from the filtered data set and analyzed by a reliable statistical procedure that includes logarithmic transformation, statistical regression, and post regression analysis. It contains UFP as the only predictor and excludes VAF, a parameter receiving much criticism. [1] The equation is flexible to the extent that it does not include any special ISBSG parameters, and thus, it can estimate FP-oriented projects and can be extended to include cost drivers for future works.

*3.5. Neural Network Learning Model*

The neural network technique is used in the processing layer of the neuro-fuzzy FP model to learn the weight values of unadjusted FP and to calibrate FP so that it reflects the trend of current software. The weight values obtained from learning via the neural network are then utilized in the adjusted fuzzy complexity weight system.

*3.5.1. Network Structure*

The neural network used in the neuro-fuzzy FP model is a typical multi-layer feed-forward network whose structure is depicted in Figure 5. The network consists of three layers: input, middle, and output.

The input layer is composed of 16 neurons, denoted as $X_i$, with $i$ ranging in value from 1 to 16. Among this group of 16, neurons $X_1$ to $X_{15}$ represent the three complexity ratings of five unadjusted FP function components. The inputs of these 15 neurons are the numbers of their respective function components. They are denoted by codes such as NINLOW (number of low External Inputs), NFLAVG (number of average Internal Logical Files), and NQUHGH (number of high External Inquiries), and are described in detail in Table VII. These 15 neurons are all connected to neuron $Y$ in the middle layer and are associated with the weights of $w_i$, with $i$ ranging in value from 1 to 15. Neuron $X_{16}$ is a bias node with a constant input of one and is connected to neuron $Z$ in the output layer with an associated weight of $v2$, which represents the coefficient $A$ in Equation 6.

---

[1] A test was done to select VAF as the predictor and included it in the equation, but the stepwise regression process automated by SPSS discarded VAF from the equation due to its poor performance.

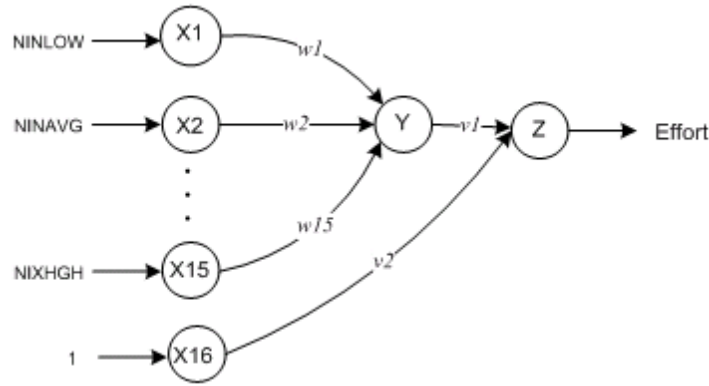

**Figure 5** - Neural Network Structure of Neuro-Fuzzy FP Model

**Table VII - NOTATION OF NEURON $X_i$ INPUTS**

|  | Low | Average | High |
|---|---|---|---|
| **External Inputs** | NINLOW $X1$ | NINAVG $X2$ | NINHGH $X3$ |
| **External Outputs** | NOULOW $X4$ | NOUAVG $X5$ | NOUHGH $X6$ |
| **External Inquiries** | NQULOW $X7$ | NQUAVG $X8$ | NQUHGH $X9$ |
| **Internal Logical Files** | NFLLOW $X10$ | NFLAVG $X11$ | NFLHGH $X12$ |
| **External Interface Files** | NIXLOW $X13$ | NIXAVG $X14$ | NIXHGH $X15$ |

In the middle layer, neuron Y receives the outputs from the 15 neurons in the input layer and is then connected to neuron Z in the output layer. The output of neuron *Y* can be expressed as: $Y = \sum_{i=1}^{15} X_i \cdot W_i$ which is functionally equivalent to the UFP calculation formula ($UFP = \sum_{i=1}^{5} \sum_{j=1}^{3} Z_{ij} \cdot W_{ij}$).

The output layer has only one neuron, *Z*, that receives the outputs from neuron *Y* of the middle layer and neuron $X_{16}$ of the input layer. The activation function of neuron *Z* is $Z = v2 \cdot Y^{v1}$ which is of the same form as Equation 6 (*Effort* = $A \cdot UFP^B$) extracted by regression analysis. Thus, neuron *Z* can be used to estimate the software cost in work effort from software size in UFP.

The underlying reason for choosing work effort as the output to train the UFP weight values is because these weight values are supposed to reflect the software component complexity. This complexity should be proportional to the project work effort, which is based on the common sense notion that the more complex the software, the more effort should be put in. Overall, the equation obtained by statistical analysis has a sound mathematical ground and a solid explanation. It is used as the activation function in the neural network, and thus, the infamous problem of the traditional neural network behaving like a black-box is avoided.

Based on the neural network structure, a back-propagation learning algorithm is conducted in order to obtain the calibrated weight values of UFP. The purpose of this algorithm is to minimize the prediction difference between the estimated and actual efforts. Given *NN* projects, the prediction difference can be expressed as the error signal defined in Equation 7:

$$E = \sum_{n=1}^{NN} \frac{1}{2}\left[\frac{Zn - Zdn}{Zdn}\right]^2 \qquad \text{Equation 7}$$

where *E* is the error signal, *Zn* is the estimated effort of the nth project and *Zdn* is the actual effort of the *nth* project, the desired output. The learning procedure is subject to monotonic constraints; in other words, the UFP weight values must be *Low < Average < High*.

### 3.5.2. Remarks on Neural Network Learning

The neural network part of the model is designed to calibrate UFP weight values and to solve the three problems with the FP complexity weight system mentioned in Section 2, which include Problem 3 (weight values obsolete), Problem 4 (weight values defined subjectively), and Problem 5 (weight values defined locally). The new calibrated weight values overcome Problems 4 and 5, because they are acquired from the ISBSG Data Repository - release 8, which is compiled from dozens of countries and covers a broad range of software industry. Furthermore, Problem 3 is also addressed because 75% of the projects are fewer than five years old.

### 3.6 Post-Tuned Fuzzy Weight Measurement System

The calibrated weight values obtained from the neural network learning are imported in the adjusted fuzzy weight measurement system and are specifically applied to the output membership functions. An example of the adjusted output membership functions of External Inputs is given in Figure 6. The adjusted weight measurement system can be used to count FP for new projects more accurately. Also, the pre-defined and adjusted fuzzy measurements constitute the complete fuzzy measurement system of the neuro-fuzzy FP model.

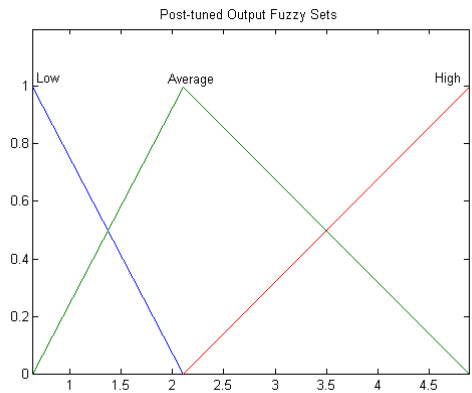

**Figure 6 - Post-tuned Fuzzy Sets for External Inputs**

## 4. MODEL ASSESSMENT

Five experiments were conducted to validate the model. For each experiment, the original data set (184 projects) was randomly separated into 100 training data points and 84 test data points. The outliers are the abnormal project data points with large noise that may distort the training result. Thus, we used the training data set excluding the outliers to calibrate UFP weight values, but used the rest of the data points to test the model.

The calibrated UFP weight values obtained from five experiments are listed in Table VIII, and the original weight values as comparison. The observation of lower weight values after calibration means that fewer work efforts are needed to accomplish the same complex software component. This is in accordance with the fact that overall productivity of the software industry has been continuously increasing since Function Points was invented in 1979.

**Table - VIII Calibrated UFP Weight Values**

| Component | Low | | Average | | High | |
|---|---|---|---|---|---|---|
| | Original | Calibrated | Original | Calibrated | Original | Calibrated |
| External Inquiries | 3 | 1.8 | 4 | 2.9 | 6 | 5.4 |
| External Outputs | 4 | 3.3 | 5 | 3.3 | 7 | 6.2 |
| External Inquiries | 3 | 1.8 | 4 | 2.9 | 6 | 5.4 |
| Internal Logical Files | 7 | 5.4 | 10 | 9.8 | 15 | 14.9 |
| External Interface Files | 5 | 4.6 | 7 | 6.9 | 10 | 10 |

The validation results of the five experiments are assessed by Mean Magnitude Relative Error (MMRE) for estimation accuracy. MMRE is defined as: for $n$ projects, $MMRE = \frac{1}{n} \sum_{i=1}^{n} \left( | Estimated_i - Actual_i | / Actual_i \right)$. The results are listed in Table IX where "Improvement %." is the MMRE improvement in percentage for each experiment. Based on the MMRE assessment results, an average of 22% cost estimation improvement has been achieved with the Neuro-Fuzzy Function Points Calibration model. The MMRE after calibration is around 100%, which is still relatively large and is due to the absence of well-defined cost drivers like COCOMO factors. Unfortunately ISBSG Release 8 does not have data on cost drivers.

**Table IX - MMRE Validation Result**

| | Exp.1 | Exp.2 | Exp.3 | Exp.4 | Exp.5 |
|---|---|---|---|---|---|
| **MMRE Original** | 1.38 | 1.58 | 1.57 | 1.39 | 1.42 |
| **MMRE Calibrated** | 1.10 | 1.28 | 1.17 | 1.03 | 1.11 |
| **Improvement %** | 20% | 19% | 25% | 26% | 22% |
| **Average Improvement %** | 22% | | | | |

The validation results of the five experiments are also assessed by Prediction at level $p$ (PRED) criteria, i.e. $PRED(p) = k / N$, where $N$ is the total number of projects, $k$ is the number of projects with absolute relative error of $p$. Four PRED criteria are assessed in this work, namely Pred 25, Pred 50, Pred 75 and Pred 100. Table X lists the PRED assessment result.

**Table X - PRED Validation Results**

|  | Average Original | Average Calibrated | Average Improvement |
|---|---|---|---|
| **Pred 25** | 13% | 12% | 0% |
| **Pred 50** | 23% | 27% | 4% |
| **Pred 75** | 40% | 46% | 6% |
| **Pred 100** | 60% | 67% | 8% |

Figure 7 plots the comparison of the original and the calibrated PRED results where the overall improvement is observed: the line with square signs (calibrated) is above the line with diamond signs (original).

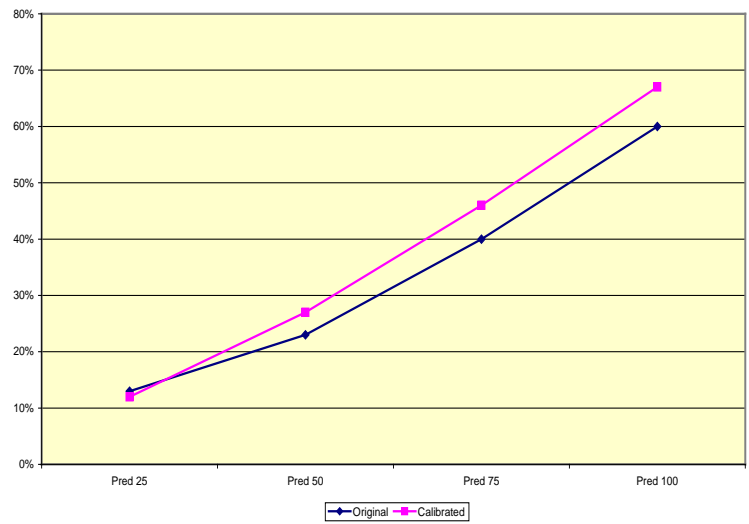

**Figure 7 - PRED Validation Results Comparison**

*4.1 Weakness of the Model*

Threats to validity are conditions that limit the researcher's ability to generalize the results of the experiment to industrial practice, which was the case with this study. Specific measures were taken to support validity; for example, a random sampling technique was used to draw samples from the population in order to conduct experiments, and filtering was applied to the ISBSG data set. Five experiments were conducted by drawing five different random samples in order to generalize the results. "Post hoc" analysis of effect size and power reinforced the validity of the experiments by yielding a large effect size.

The proposed calibration of the FP element's weights were applied to the ISBSG data set to monitor the effectiveness of the approach; a potential threat to the validity of this study involved the question of whether or not similar results would be obtained with an entirely different sample. In this investigation, we calibrated the weights of the five FP elements using only the ISBSG data set, which has raised a threat to the validity of the calibration process. The ISBSG data set contains projects using different function point counting techniques, such as IFPUG, COSMIC and MARK II. Because 90% of the sample used the IFPUG counting method, we therefore restricted our experiments to IFPUG projects. This decision may lead to the question as to whether the proposed model's outcome will be valid if the model is used with the other two types of FP counting technique besides IFPUG.

ISBSG - release 8 is a large and wide-range project data repository, so the calibrated FP weight values learned from this repository reflect the maturity level of the software industry at that time. However, software development is a rapidly growing industry and these calibrated weight values will not reflect tomorrow's software. In the future, when modern project data is available, the FP weight values will again need to be re-calibrated to reflect the latest software industry trend.

Our study is a data-driven type of research where we extracted a model based on known facts. The proposed model is more meaningful for small projects, which are actually the most common type of projects in the software industry. This limitation is due to the ISBSG data set characteristics used in this study, and it may raise concerns about the validity, specifically with large projects. In reality, there are more small projects than large ones, and even the large projects tend to be subdivided into smaller projects, so that they become easier to manage. Although the proposed approach has some potential to threaten the model's validity, we followed appropriate research procedures by conducting and reporting tests to guarantee the reliability of the study, and certain measures were also taken to ensure its validity.

## 5. CONCLUSIONS

FP as a software size metric is an important topic in the software engineering domain. The validation results of the neuro-fuzzy FP model with the empirical data repository (ISBSG - release 8) show a 22% improvement in software cost estimation. This result indicates that the original unadjusted FP weight values require updated calibration for more accurate cost estimations. This paper provides a framework to calibrate the complexity weight values and solves the problems with the FP weight mentioned in Section 2.

The fuzzy part of the neuro-fuzzy FP model produces an exact complexity degree for each functional component by processing its associated file numbers using fuzzy logic. This part of the model overcomes two problems with the unadjusted FP complexity classification: ambiguous classification (Problem 1) and crisp boundary (Problem 2), as described in sub-section 3.3.

The neural network part of the neuro-fuzzy FP model calibrates UFP weight values using the ISBSG Data Repository - release 8, which contains 2,027 projects from dozens of countries and covers a broad range of project types from many industries, with 75% of the projects being less than five years old. This part of the model overcomes three problems with the unadjusted FP complexity weight values: obsolete weight values (Problem 3), weight values defined subjectively (Problem 4), and weight values defined locally (Problem 5) as laid out in sub-sections 3.5 and 3.6.

Finally, sub-section 3.4 presents a new equation to estimate the software cost in work effort initially derived from the ISBSG Data Repository - release 8. It is further improved with the reliable filtered data set and analyzed by a reliable statistical procedure. This equation fulfills the requirement of being flexible and integral, and has the potential to involve other cost drivers in future research.